\documentclass[runningheads]{llncs}
\usepackage{amssymb}
\usepackage{amsmath}
\usepackage{listings}
\usepackage{xcolor}

\newif\ifsubmit
\submitfalse
\ifsubmit

\newcommand{\MDP}[1]{#1}
\newcommand{\MDPComm}[1]{}
\newcommand{\Elena}[1]{} 
\newcommand{\Silvia}[1]{} 

\else

\newcommand{\MDP}[1]{\textcolor{red}{#1}}
\newcommand{\MDPComm}[1]{{\scriptsize \MDP{[Matteo{:} #1]}}}
\newcommand{\Elena}[1]{\mnote{\textbf{\color{blue}[EZ: #1]}}} 
\newcommand{\Silvia}[1]{\mnote{\textbf{\color{magenta}[SC: #1]}}} 
\fi

\newcommand{\olds}[1]{\oldstylenums{#1}}
\newcommand{\oldsb}[1]{{\bfseries\olds{#1}}}
\newcommand{\mnote}[1]{\stepcounter{ncomm}%
\vbox to0pt{\vss\llap{\scriptsize\oldsb{\arabic{ncomm}}}\vskip6pt}%
\marginpar{\scriptsize\bf\raggedright%
{\oldsb{\arabic{ncomm}}}.\hskip0.5em#1}}
\newcounter{ncomm}

\newcommand{\address}{\texttt{address}}
\newcommand{\addressPay}{\texttt{address payable}}
\newcommand{\msgsender}{\texttt{msg.sender}}
\newcommand{\this}{\texttt{this}}
\newcommand{\fb}{\emph{fallback}}
\newcommand{\transfer}{\texttt{transfer}}
\newcommand{\AddressType}[1]{\texttt{address}\langle#1\rangle}
\newcommand{\Topfb}{\texttt{Top}_{\texttt{fb}}}
\newcommand{\Top}{\texttt{Top}}

\lstdefinelanguage{FJS}{
	keywords={contract, this, return, uint, unit, mapping, address, msg, sender, value, u, if, then, else},
}

\lstdefinestyle{basiclisting}{
	 mathescape=true,
	basicstyle=\ttfamily\small,%\scriptsize,
	keywordstyle=\bfseries,%\color{darkRed},
	numbers=none,
	stepnumber=1,
	escapechar=|
}

\begin{document}
\title{%The gap between typed and type-safe is...formal methods!\\
Solidity 0.5: when typed does not mean type safe
%\thanks{Supported by organization x.}
}
\author{Silvia Crafa\inst{1}\and Matteo Di Pirro\inst{2}}

%\authorrunning{S. Crafa et al.}
\institute{University of Padova, Italy \quad
 \email{crafa@math.unipd.it} \and
Kynetics, Italy\quad
\email{matteo.dipirro@kynetics.com}
 }
\maketitle              % typeset the header of the contribution
\begin{abstract}
The recent release of Solidity 0.5 introduced a new
type to prevent Ether transfers to smart contracts that are 
not supposed to receive money. Unfortunately, the compiler
fails in enforcing the guarantees this type intended to
convey, hence the type soundness of Solidity~0.5 is no better than
that of Solidity~0.4.
In this paper we discuss a paradigmatic example showing that
vulnerable Solidity patterns based on potentially unsafe callback
expressions are still unchecked. 
We also point out a solution that
%firmly rooted in 
%well foundend on 
strongly relies on formal methods to support a type-safer smart
contracts programming discipline, while
being retro-compatible with legacy Solidity code. 
\end{abstract}

\setcounter{footnote}{0}

\section{Introduction}

Over %\Silvia{dici di piu' su che type soundness Solidity 4}
 the last few years the execution of smart 
contracts on the blockchain has emerged as a
form of distributed programming of a global computer. Anyone can
deploy a global service, encoded as a smart contract, that can be used
by mutually untrusted parties to ``safely'' interact with no need of a
central authority. Therefore it is of paramount importance that the
intended interaction provided by the service is ``correctly''
implemented by the code of the corresponding contract. 
Indeed, while the term \emph{contract} is generally used to 
refer to an interaction that is intended to be enforced by law,  
a smart contract on the blockchain is intended to be
\emph{automatically enforced}: the law is embodied by the code to be
executed (see the TheDAO affair~\cite{theDao}).
%once a contract
%is deployed with bugs there is no way to correct them, no matter how
%serious the consequences are.
%

Formal methods 
%can be useful also in the new context of blockchains. As a matter of
%fact, they 
have a long tradition of successes in dealing with the
subtle mismatches between program specification 
% /service/contract 
and code implementation, and they can be helpful also in the new
context of smart contracts.
% In the case of Ethereum, there are
%two levels of the latter: smart contracts are written in a
%high level programming language, where Solidity is the most popular
%and most used, and are then compiled into bytecode that runs on the
%Ethereum Virtual Machine. 
Here we focus on Solidity, the most widely used programming language
in Ethereum's ecosystem, and on formal methods that provide support
for a safer programming discipline by acting directly at the
programming language  level.  
In particular, since Solidity is a statically typed language, we
%resort to the type theory of high level languages, 
foster the use of types as a tool to shape and
substantiate the programmer's reasoning.  
%First of all, we observe that Solidity contracts are
%reminiscent of class-based objects in distributed OOLs. Therefore 
%it is worth to study how the rich and well-known theory of OOLs can
%be reused and adapted to smart contracts. 
%Secondly, 
%Moreover, experience  has shown that 
However, static typing conveys an effective programming discipline
only if type constraints are actually enforced by the compiler. 
In other terms, there is a gap between the definition
of types in a language and their type-safe usage.
We show below that this is precisely the case of
the last release of Solidity 0.5. 
Indeed, the newly-introduced type $\addressPay$ is intended to prevent
Ether transfers to smart contracts that are not supposed to receive
money, but the compiler fails to enforce such semantics. %, thus
In other words, the type
soundness of Solidity 0.5 is no better than that of the previous
release.

Formal methods and the theory of typed languages show the way to
bridge that gap and develop a statically typed language that is also
type-safe. In particular, since Solidity contracts are
reminiscent of class-based objects in distributed Object-Oriented
Languages, it is worth to study how the rich and well-known theory of
OOLs can be reused and adapted to smart contracts programming.

%In other words, well-typed programs do not dynamically evolve to an
%error state caused by a violation of the safety property that types
%intended to guarantee.\Silvia{mi sa che la facciamo troppo lunga}
%We show below that Solidity 0.5 recently introduced a
%\emph{type-unsafe} access to smart contracts. 
%We discuss the motivations and the
%consequences of this issue, and 
In a previous work 
we defined the Featherweight Solidity typed calculus (\cite{wtsc19}),
which %we precisely formalized 
formalizes the core of the Solidity language and the basic 
type soundness provided by its compiler (both versions 0.4 and 0.5). 
%and the behavior of smart
%contracts. The type system of Featherweight Solidity 
%and Solidity's compiler in terms of the operational semantics and the
%type system of a core calculus (Featherweight Solidity). 
In that work we also proposed a refined typing that enjoys a stronger
soundness property, but remains retro-compatible with legacy Solidity
code. That typing ensures safer accesses to contracts through
their address; hence it statically prevents a general class of runtime
errors.
We show here that the unsafe usage of the $\addressPay$ type
can be statically captured by the refined type system put forward in
\cite{wtsc19}. %That typing then
Therefore, it represents a solution to the
soundness issue of Solidity 0.5 and supports an effective %a safer
smart contract programming discipline using the compiler as a convenient building tool.
% corresponding to vulnerable Solidity programming patterns.  
%  avoids the unsafe usage of the
%$\addressPay$ type by providing a solution that statically prevents a
%more general class of runtime errors. 
%, thus promoting
%a safer smart contracts programming discipline in Solidity.

\section{The problem}

As in class-based Object-Oriented Languages, % (OOLs), 
the declaration of
a Solidity contract $C$ defines a contract type $C$. However,
%In Ethereum the instances of smart contracts deployed on the
%blockchian can only be accessed through their public address, which
%essentially represents an untyped way to access such instances. 
%Accordingly, given a Solidity contract $C$, 
instances of such a contract are often referred to by the Solidity
code through expressions of type $\address$, that essentially
represent an untyped way to access them. Such expressions must then be
cast to the type $C$ in order to call the functions
provided by the contract $C$.    
%Exploiting the analogy with class-based OOLS, the type $\address$
%corresponds to an untyped pointer that is cast to a class type
%so to invoke the methods provided by that class. 
Casting an untyped pointer is notoriously a very flexible but subtle
feature requiring programmers to precisely know what pointers
refer to.
%, and the compiler to implement some form of static pointer analysis.
Solidity's compiler provides no help here: neither static or dynamic
checks 
are performed on cast expressions, and a dynamic error is raised only
when calling a function (or accessing a state variable) that is not
provided by the underlying contract.

Two features of Solidity make this problem pervasive in the code of
smart contracts. First of all, in Ethereum the instances of smart
contracts deployed on the blockchain can only be accessed through
their public address. 
%contracts can refer to other contract instances only through their
%public addresses. 
Secondly, contract functions make extensive use of their implicit
variables $\this$ and $\msgsender$, that are dynamically bound to the
\emph{contract instance} being executed and the 
\emph{address} of the caller contract, respectively. Therefore, while
the callee is referred to through a typed pointer (as in OOLs), the
caller is referred to through an untyped one.
Hence, even though usual method recursion is type-safe, all the callback
expressions undergo potentially unsafe usages. %casts. 
Indeed, besides the dangerous casts described above, 
a typical Solidity pattern consists in calling
$\msgsender.\transfer(n)$
to send $n$ Ether from the balance of the callee to that of the
caller. However, such a transfer implicitly calls the $\fb$ function
of the contract referred to by $\msgsender$, thus raising a dynamic error
if such function has not been defined by that contract.

To mitigate this problem, the last release of Solidity (i.e. version
0.5~\cite{solidity5}) distinguishes two types, $\address$ and
$\addressPay$, where the second one denotes addresses pointing
to contracts that declare the $\fb$ function. Ideally, by using the new type
$\addressPay$, Solidity 0.5 intends to statically prevent at least
the unsafe money transfers, that are actually the most common
form of the dynamic errors described above.
It is worth to observe that these errors, that in OOLs are known as
\emph{message-not-understood}, are particularly harmful in the
context of the blockchain. Indeed, in Ethereum the
occurrence of a dynamic error causes the initial transaction to be
interrupted and rolled-back (the so-called \textit{revert}).
This makes the account that issued that
transaction lose the money it paid to the miner node and possibly
leads to Ether indefinitely locked into a contract's balance.
Hence, there is a pressing requirement to issue a transaction only if
it can be statically guaranteed that it will not evolve to
a revert. 

Unfortunately, Solidity 0.5 fails to prevent unsafe
money transfers at compile-	time. As a matter of fact, no type
check is enforced by the
compiler to ensure that a variable of type $\addressPay$ is
substituted with the address of a contract that actually provides a
$\fb$ function. The problem can be detected with 
a careful read of the documentation\footnote{
   \url{https://solidity.readthedocs.io/en/v0.5.9/050-breaking-changes.html}
}, which states:
\begin{quote}
\emph{
It might very well be that you do not need to care about the
distinction between address and address payable and just use address
everywhere. For example, if you use the withdraw pattern you most
likely do not have to change your code because transfer is only used
on msg.sender instead of stored addresses and msg.sender is an address
payable.\\ 
\mbox{[...]}
Address literals can be implicitly converted to address payable. \\ 
\mbox{[...]}
In external function signatures address is used for both the address
and the address payable type. 
}
\end{quote} 

\begin{figure}[h!]
{\small
\begin{lstlisting}[style=basiclisting,numbers=left,numberstyle=\tiny\color{gray}]
pragma solidity >=0.5.0 <0.7.0;

contract WithoutFallback {
  Test _test;
    
  constructor (address _unsafeAddress) payable public {
      _test = Test(_unsafeAddress);
  }    

  function callUnsafeContract() external {
      _test.foo();
  }    

  function testUnsafeCast() external {
     address _addr = address(_test);
     //_addr.transfer(10); // DOES NOT COMPILE
     address payable _payAddr = address(uint160(_addr));
     _payAddr.transfer(10);
    }
}

contract Test {
  constructor () payable public {}
    
  function foo() external {
      msg.sender.transfer(10);
  }    
}
\end{lstlisting}
}
\caption{Counterexample to the type safety of Solidity 0.5}
\label{fig:counterEx}
\end{figure}

Concretely, the counterexample in Figure~\ref{fig:counterEx} 
shows that the implicit
variable $\msgsender$ is assumed to be of type $\addressPay$,
but no check is performed on the type of the actual caller's address.
More precisely, the expression $\msgsender.\transfer(10)$ in the body
of the function of the contract \texttt{Test} (line 26)
correctly compiles, and so does the call of this function from the
contract \texttt{WithoutFallback} (line 11). 
However, issuing a transaction that invokes
the function \texttt{callUnsafeContract} of \texttt{WithoutFallback}
results in a revert as 
that contract cannot receive money back from the contract
\texttt{Test}. The same problem occurs if the functions are marked
public or private instead of external.  
Furthermore, in order for the 
contract \texttt{WithoutFallback} to refer to a deployed instance of
the contract \texttt{Test}, its constructor can only accept a
parameter of address type %sia \address che \addressPay
 and then cast it to the expected contract
type (line 7). Even if nothing ensures that the actual parameter
refers to an instance of \texttt{Test}, the cast expression correctly
compiles 
and correctly executes, postponing the dynamic check to the moment
where the \texttt{\_test} reference is actually used (line 11). 
The constructor's parameter \texttt{\_unsafeAddress} could also be of
type $\addressPay$. In this case, one might expect the compiler to
check that, when casting a payable address to a contract type, 
the target type of the cast (i.e. \texttt{Test}) at least
defines a $\fb$ function. Again, this is not true. No check is
performed, either at compile-time or at run-time, to ensure that
\texttt{Test} respects the constraints that $\addressPay$ is supposed
to impose. 

The example also shows (in function \texttt{testUnsafeCast}) that the
$\transfer$ primitive can be correctly used only on addresses of
static type $\addressPay$, but the type constraint can be circumvented
by resorting to an intermediate cast to the type \texttt{uint160}, as
explicitly stated by the official  
documentation. Clearly, the expression
\texttt{\_payAddr.transfer(10)} at line 18 dynamically raises an
error since there is no $\fb$ function in the \texttt{Test}
contract. 

We tested the code in Figure \ref{fig:counterEx} with Remix,
the online Ethereum IDE, using the version
\textit{0.5.9+commit.e560f70d} 
of the Solidity compiler.

Money transfers that dynamically lead to errors were possible since
the first release of Solidity, so the new version has not introduced
a new problem. On the other hand, the addition of the new type
$\addressPay$ to capture the (addresses of) contracts that can
``safely'' receive Ether, generates into programmers the expectation
that ``safely'' means type-safely, that is the compiler will check
it. In fact nothing has actually changed w.r.t. version 0.4: the
new type essentially provides only a refined documentation about 
addresses, but programmers have certainly more confounded
expectations. 

\section{The solution}

%Experience with formal methods and the typed theory of programming
% languages allows to identify a type preservation issue form the
% Solidity documentation quoted above. The solution comes again from
% the toolbox of formal methods
The typed theory of programming languages allows to identify a type
preservation issue in Solidity 0.5's type system, confirmed 
by the code in Figure \ref{fig:counterEx}, and also offers a
solution. %\Silvia{vista la venue metti accento sul valore dei FM}
%, as confirmed 
%by the code in Figure \ref{fig:counterEx}, Solidity 0.5 contains
%a type preservation issue.
In a previous work (\cite{wtsc19}) we developed a precise
formalization of the core of the Solidity language and its type
system. We resorted to a formalization style that is reminiscent
of the well known Featherweight Java language~\cite{fj}, highlighting the
similarities between the notions of object and smart contract. Along with
a precise definition of the basic type-soundness provided by
the Solidity compiler, we proposed a refined type system that enjoys a
stronger soundness property. In particular, that typing solves
the type preservation problem pointed out here. 
Furthermore,
the solution put forward in~\cite{wtsc19} is general enough to
statically prevent not just unsafe 
calls to a non existent $\fb$ function, but all the
\emph{message-not-understood} errors arising from unsafe casts from
addresses to contract types.
%cast expressions or money transfers that would lead to unsafe usage
%of contract members or calls to an undefined fallback function are
%now ruled out at compile-time 

The key idea is twofold. First, the type $\address$ is refined with 
type information about the contract it refers to. That is, 
$\AddressType{C}$ is the type of the addresses of instances of the
contract $C$, or of a contract that inherits form $C$.  
In particular, assuming a dummy contract $\Topfb$ that only contains 
a $\fb$ function with an empty body, the type $\AddressType{\Topfb}$
has the same meaning of Solidity 0.5's $\addressPay$. Indeed,
it is the (super-)type of the addresses of every contract that can
safely accept money transfers.\footnote{In \cite{wtsc19} we proposed 
  the keyword \texttt{payableaddress} as a syntactic sugar for the
  type $\AddressType{\Topfb}$, since at the time of writing we were
  not aware of Solidity 0.5.}

The second idea is to enrich %essential to guarantee type preservation
functions' signatures with the maximum type
allowed for the caller, so that functions can only be invoked by
contracts with an expected (super-)type.
%whose address points to a contract with the expected (super-)type. 
%we enrich functions' signatures with the address type of the
%implicit sender parameter, so to allow functions to be
%called only by contracts whose address has an expected (super-)type. 
Adding a type constraint for the caller in function
signatures is essential to safely type the implicit $\msgsender$
parameter, thus to guarantee type preservation. 
The compiler can then statically check potentially unsafe
callback expressions, such as $\msgsender.\transfer(n)$ or
\\\texttt{C(msg.sender).foo()}, that reduce to a revert 
if $\msgsender$ is bound to the address of a contract that
has no $\fb$ function or does not have type $C$,
respectively.

The counterexample in Figure \ref{fig:counterEx} can be fixed 
by choosing a suitable refined signature for the \texttt{foo}
function of the contract \texttt{Test}. As the only requirement for
the caller is to provide a $\fb$ function, it is
sufficient to amend the function's code as follows:
\begin{lstlisting}
function foo() <$\Topfb$> external {
   msg.sender.transfer(10);
}
\end{lstlisting}
In the body of the function, the variable $\msgsender$ is then assumed
to have type $\AddressType{\Topfb}$, hence the call to
\texttt{transfer} is now well typed. On the other hand, 
the compiler prevents the unsafe money transfer by 
identifying a type error in the function call at line 11, since the
caller's type, \texttt{WithoutFallback}, is not a subtype of 
$\Topfb$.  
As a further example, the following function, whose refined signature
specifies the expected (super-)type of the caller, could
be safely added to the \texttt{Test} contract:
\begin{lstlisting}
function boo() <WithoutFallback> external {
   WithoutFallback(msg.sender).testUnsafeCast();
}
\end{lstlisting}
To simplify the notation, and in line with the Solidity programming
style, in~\cite{wtsc19} we proposed a syntactic sugar based on a
new function marker, 
\texttt{payback}, for functions whose caller must simply
provide a $\fb$ function (which is the most common case). 
In this way the \texttt{foo} function inside the \texttt{Test}
contract would simply become as follows:
\begin{lstlisting}
function foo() payback external { 
   msg.sender.transfer(10); 
}
\end{lstlisting}
Similarly, the standard function signature with no annotation 
could correspond to assuming the (super-)type $\AddressType{\Top}$,
that is no constraint for the caller. 
%Note that this refinement is not always mandatory. If a function
%does not need to impose any constraints on its callers, no annotation
%is required.

Further details about the formalization of this idea, its
type-soundness, and its retro-compatibility with Solidity 
contracts already deployed on the blockchain can be found
in~\cite{wtsc19}. 
We just observe here that, despite the usage of the convenient
\texttt{payback} marker, to take advantage of the full 
power of the refined typing the major effort required to Solidity
programmers is to annotate their functions with the required (super-)type
of the caller. Such a requirement might be verbose, but it actually
supports a safer programming discipline, where types mirror the
programmer's reasoning and the compiler can be effectively used as a
convenient building tool.

%\bibliographystyle{abbrv}
%\bibliography{main}

\end{document}